\documentclass[aps,prd,superscriptaddress,preprintnumbers,nofootinbib,10pt]{revtex4-2}
\usepackage{braket}
\usepackage{multirow}
\usepackage{amsmath}
\usepackage{amssymb}
\usepackage[dvipdf,dvips]{graphicx}
\usepackage{color}
\usepackage{hyperref}
\usepackage{url}
\usepackage{slashed}
\usepackage{subfigure}
\usepackage[usenames,dvipsnames]{xcolor}
\usepackage{amsmath}
\usepackage{amsfonts}
\usepackage{float} 
\usepackage{amssymb}
\usepackage{epsfig}
\usepackage{graphics}
\usepackage{euscript}
\usepackage{slashed}
\usepackage{epstopdf}
\usepackage[utf8]{inputenc}
\allowdisplaybreaks
\usepackage[normalem]{ulem}
\usepackage{pifont}
\usepackage{dsfont}
\usepackage{MnSymbol}
\usepackage{graphicx}
\usepackage{latexsym}
\usepackage{tikz-feynman}
\usepackage{tikz-cd}
\usepackage{easyReview}
\usepackage{cancel}
\usepackage[normalem]{ulem}
\newcommand\beq{\begin{eqnarray}}
\newcommand\eeq{\end{eqnarray}}

\counterwithin{table}{section}
\counterwithin{figure}{section}

\makeatletter
\DeclareRobustCommand{\cev}[1]{%
	{\mathpalette\do@cev{#1}}%
}
\newcommand{\do@cev}[2]{%
	\vbox{\offinterlineskip
		\sbox\z@{$\m@th#1 x$}%
		\ialign{##\cr
			\hidewidth\reflectbox{$\m@th#1\vec{}\mkern4mu$}\hidewidth\cr
			\noalign{\kern-\ht\z@}
			$\m@th#1#2$\cr
		}%
	}%
}

\makeatother
\begin{document}

\title{\bf The setting sun diagram with complex external momenta}

\author{D.~Dudal}
\email{david.dudal@kuleuven.be} 
\affiliation{KU Leuven Campus Kortrijk–Kulak, Department of Physics,	Etienne Sabbelaan 53 bus 7657, 8500 Kortrijk, Belgium and Ghent University, Department of Physics and Astronomy, Krijgslaan 281-S9, 9000 Gent, Belgium}

\author{D.M.~van Egmond}
\email{duifjemaria@gmail.com} 
\affiliation{Instituto de F\'isica Te\'orica, Universidade Estadual Paulista, Rua Dr. Bento Teobaldo Ferraz, 271-Bloco II, 01140-070 S\~ao Paulo, SP, Brazil
}
\affiliation{ICTP South American Institute for Fundamental Research Instituto de F\'isica Te\'orica, UNESP - Univ. Estadual Paulista, Rua Dr. Bento Teobaldo Ferraz, 271, 01140-070 S\~ao Paulo, SP, Brazil}

\author{G.~Krein}
\email{gastao.krein@unesp.br}
\affiliation{Instituto de F\'isica Te\'orica, Universidade Estadual Paulista, Rua Dr. Bento Teobaldo Ferraz, 271-Bloco II, 01140-070 S\~ao Paulo, SP, Brazil
}

\begin{abstract}
We revisit the issue of analytically continuing Feynman integrals from Euclidean to Minkowski signature, allowing for generic complex momenta. Although this is well-known in terms of the K\"all\'{e}n-Lehmann representation, we consider potential alternative takes on the same problem and discuss how these are not necessarily equivalent to the K\"all\'en-Lehmann integral outcome. We present our analysis for a simple enough case---the setting sun diagram in $d=2$ with a real mass---but already with an eye out to the more general case with complex masses which will further complicate matters.
\end{abstract}
\maketitle
\section{Introduction}\label{sectintro}

Complex pole masses of quarks and gluons, or more generally speaking their Green's functions analytic structure in the complex momentum plane, appear in various methods for non-perturbative QCD in Euclidean spacetime, from analytical \cite{Gribov:1977wm,Zwanziger:1988jt,Krein:1990sf,Alkofer:2003jj,Baulieu:2009ha,Windisch:2012sz,Windisch:2012zd,Canfora:2013zna,Huber:2018ned,Huber:2022nzs,Fischer:2020xnb,Hayashi:2020few,Hayashi:2021jju,Cyrol:2018xeq,Siringo:2022nok,Siringo:2022dzm,Li:2019hyv,Li:2021wol,deBrito:2024ffa} and semi-analytical \cite{Asorey:2024mkb,Buoninfante:2025klm,Anselmi:2025uzj} to numerical\cite{Cucchieri:2011ig,Dudal:2018cli,Tripolt:2018xeo,Binosi:2019ecz,Lechien:2022ieg,Boito:2022rad}. At the same time, a Minkowskian description of this region is much less understood, even though it is a prerequisite to understand particles in the real (Minkowski metric) world, including their scattering and decaying properties. The purpose of this short paper is to address some curiosities about the relation between Euclidean and Minkowskian results for complex external momenta $p$. This relation is nontrivial since complex external momenta may obstruct the usual Wick rotation between Euclidean and Minkowskian spacetime, either due to poles and/or branch points and associated cuts at unusual places. Some early observations concerning this can be found in \cite{Baulieu:2009ha}.  More generally speaking, understanding Green's functions with complex external momenta will be anyhow necessary when masses of (evidently unphysical) constituent particles are or can dynamically become complex-valued.

We will concentrate on one of the simplest examples in which such issues might already arise: the two-dimensional setting sun diagram with two equal real masses $m$, which in Euclidean spacetime is written as
\beq
S_E\left(p^2\right)=\int \frac{d^{2} k_E}{(2 \pi)^2}\frac{1}{\left(k_E^2+m^2 \right)} \frac{1}{\left((k_E-p)^2+m^2 \right)},
\label{fo}
\eeq
with $k_E^2=k_2^2+k^2$. On the other hand, the Minkowskian version of the same diagram is given by
\beq
S_M\left(p^2\right)=-\int \frac{d^{2} k_M}{(2 \pi)^2}\frac{i}{\left(k_M^2-m^2+i \epsilon \right)} \frac{1}{\left((k_M+p)^2-m^2+i \epsilon \right)},
\label{fo2}
\eeq
with now $k_M^2=k_0^2-k^2$.\footnote{Here, we use a single expression for $p^2$ in both Euclidean and Minkowski space, since we can, without loss of generality, align $p$ along the $k_0$ and $k_2$ directions, respectively.} The goal of this paper is to investigate the well-known relation $S_E(p^2)=S_M(-p^2)$ for all $\textit{a priori}$ complex $p^2$, by which we mean that $p^2$ can be complex-valued before performing the integrals in \eqref{fo} and \eqref{fo2}.

For real masses $m$ one can use the K\"all\'en-Lehmann spectral representation to explore, a posteriori, the whole complex $p^2$-plane, as we will review in Section 2. This means that the usual connection between Euclidean and Minkowskian spacetime for real $p^2$ extends to the complex $p$-plane through the K\"all\'en-Lehmann spectral representation. 

As we will show in Section 3, however, surprisingly if one starts from $S_E(p^2)$ with an a priori complex $p^2$, a different relation between $S_E(p^2)$ and $S_M(p^2)$ is found. This is in contradiction with the results from the K\"all\'en-Lehmann representation, and we therefore have to conclude that starting from the integral with complex $p^2$ does not give the expected results. A possible explanation is that Green's functions in momentum space are in principle defined by taking appropriate functional derivatives of a generating functional containing momentum-dependent sources, next to the momentum-integral of the Lagrangian, that is, the action. The latter integrals are typically only written down for either Euclidean or Minkowskian momentum choice, with actions relatable to each other via the Wick rotation, assuming the rotation is permitted. To the best of our knowledge, directly obtaining a Green's function for generic complex momentum variables would require defining ab initio an appropriate generating functional, and action, also with generic complex momentum variables. It is not clear to us whethet this would be a concrete possibility, neither theoretically nor computationally. Another option would be to first compute everything in real (coordinate) space and then analytically continue, insofar possible, the necessary Fourier transformations, which usually require assumptions on the imaginary parts of the Fourier (momentum) variables to enforce convergence.

In Section 4, we will delve into yet another proposal to define the Minkowski integral, following \cite{Eichmann:2019dts}, and compare with our results. In this first study, we will stick to real $m$, so that we can use the known K\"all\'en-Lehmann results as a benchmark. However, for complex $m$ the K\"all\'en-Lehmann integral will not even be well defined, and it remains to be seen how to define a sensible relation between the Euclidean and Minkowskian integrals. To the best of our knowledge, it is not even known beforehand if this is possible. This will be dealt with in a future work, as complex poles have emerged in various effective descriptions of Green's functions in non-perturbative Yang-Mills theories, such as \cite{Dudal:2008sp,Baulieu:2009ha,Kondo:2019rpa,Siringo:2022dzm}.

\noindent

\section{The K\"all\'en-Lehmann spectral representation}\label{sectKL}
The K\"all\'en-Lehmann representation relies on the fact that for any quantum field theory, the propagator should be expressable in spectral integral form, with positive spectral density \cite{Peskin:1995ev,Coleman:2011xi}. This reveals the deep relationship between the physical states of the theory and their contributions to the correlation functions. In the case of the Euclidean two-point function in Eq. \eqref{fo}, we can follow the conventions of \cite{Dudal:2010wn} to write

\beq
S_E(p^2)= \int_{4m^2}^{\infty} d \tau \frac{\rho(\tau)}{\tau+p^2}\, \qquad \textrm{with}~\rho(\tau)= \frac{1}{2\pi}\frac{1}{\sqrt{\tau^2-4m^2 \tau}}.
\label{se}
\eeq
Based on Cauchy's formula, one can infer the the fundamental connection
\beq
\rho(\tau)=\frac{1}{2\pi i}\lim_{\epsilon\to0^+}\left( S_E(-p^2-i\epsilon)-S_E(-p^2+i\epsilon)\right)
\eeq
The threshold is given by $(m+m)^2$. Since the form \eqref{se} is analytic in the whole complex $p^2$-plane except for the interval $[-\infty,-4 m^2]$, this allows to use it as a definition of the original Feynman integral over
the whole cut complex plane. In particular, this means that the relation $S_E(p^2)=S_M(-p^2)$ extends to the complex plane, where $S_M(p^2)$ has a discontinuity in Minkowski space along the positive real axis over the interval $[4 m^2, \infty]$. This integral representation is also consistent with unitarity, under the form of the optical theorem and, closely related, the Schwartz reflection principle \cite{Peskin:1995ev,zwicky2016briefintroductiondispersionrelations}.

\section{Integral with a priori complex external momentum}\label{22}

In this Section, we will start from the Euclidean setting sun diagram in two dimensions as given in Eq.~\eqref{fo2} and investigate in detail its relation to the Minkowskian setting sun diagram in two dimensions as given in Eq.~\eqref{fo}, using the techniques of Wick rotation, paying attention to potential subtleties. We have chosen to work in $d=2$ as then we do not even need to worry about regularization/renormalization.

Setting $k_E= (k_2,k)$, we thus start from
\beq
S_E\left(p^2\right)=\frac{1}{(2 \pi)^2} \int d k \int d k_2\frac{1}{\left(k_2^2+k^2+m^2 \right)} \frac{1}{\left((k_2-\sqrt{p^2})^2+k^2+m^2 \right)}
\label{two2},
\eeq
where, without loss of generality, we have defined the external momentum $p$-vector as $\begin{pmatrix} \sqrt{p^2} \\ 0 \end{pmatrix}$. We have left the square root explicit since $\sqrt{p^2}=\pm p$ is multivalued and we have to specify a branch on the complex $p^2$-plane \cite{Lo:2020phg}. We choose $\sqrt{p^2}$ such that its real part is always positive, i.e. $\sqrt{p^2}=\sqrt{\vert p^2 \vert} e^{i \text{Arg}(p^2)/2}$ with $-\pi < \text{Arg}(p^2) < \pi$. In this way, $\sqrt{p^2}$ is well defined everywhere except on the negative real $p^2$-axis, and also $\overline{\sqrt{p^2}}=\sqrt{\overline{p^2}}$ for all $p^2 \in \mathbb{C}/\mathbb{R}_-$. \footnote{We use the $\overline{\phantom{x}}$-notation for complex conjugation.}

The $k_2$ integrand has 4 $k$-dependent poles
\beq
f_1= i \sqrt{k^2+m^2},\,\,\, f_2= -i \sqrt{k^2+m^2},\,\,\, f_3= i\sqrt{k^2+m^2}+\sqrt{p^2},\,\,\, 
f_4= -i \sqrt{k^2+m^2}+\sqrt{p^2},
\eeq
with residues of the integrand at each pole $f_x$ given by $R_x$, defined as 
\beq
R_1 (k,\sqrt{p^2})=-R_2(k,-\sqrt{p^2})=R_3 (k,-\sqrt{p^2})=-R_4 (k,\sqrt{p^2})=\frac{-i}{2 p^2} \left( \frac{1}{\sqrt{k^2+m^2}}-\frac{1}{\sqrt{k^2+m^2}+\frac{i \sqrt{p^2}}{2}}\right),
\label{ten}
\eeq
where it is clear that the residues depend explicitly on $\sqrt{p^2}$ and have a branch cut along the negative real $p^2$-axis.

\begin{figure}[!h]
	\includegraphics[width=\textwidth]{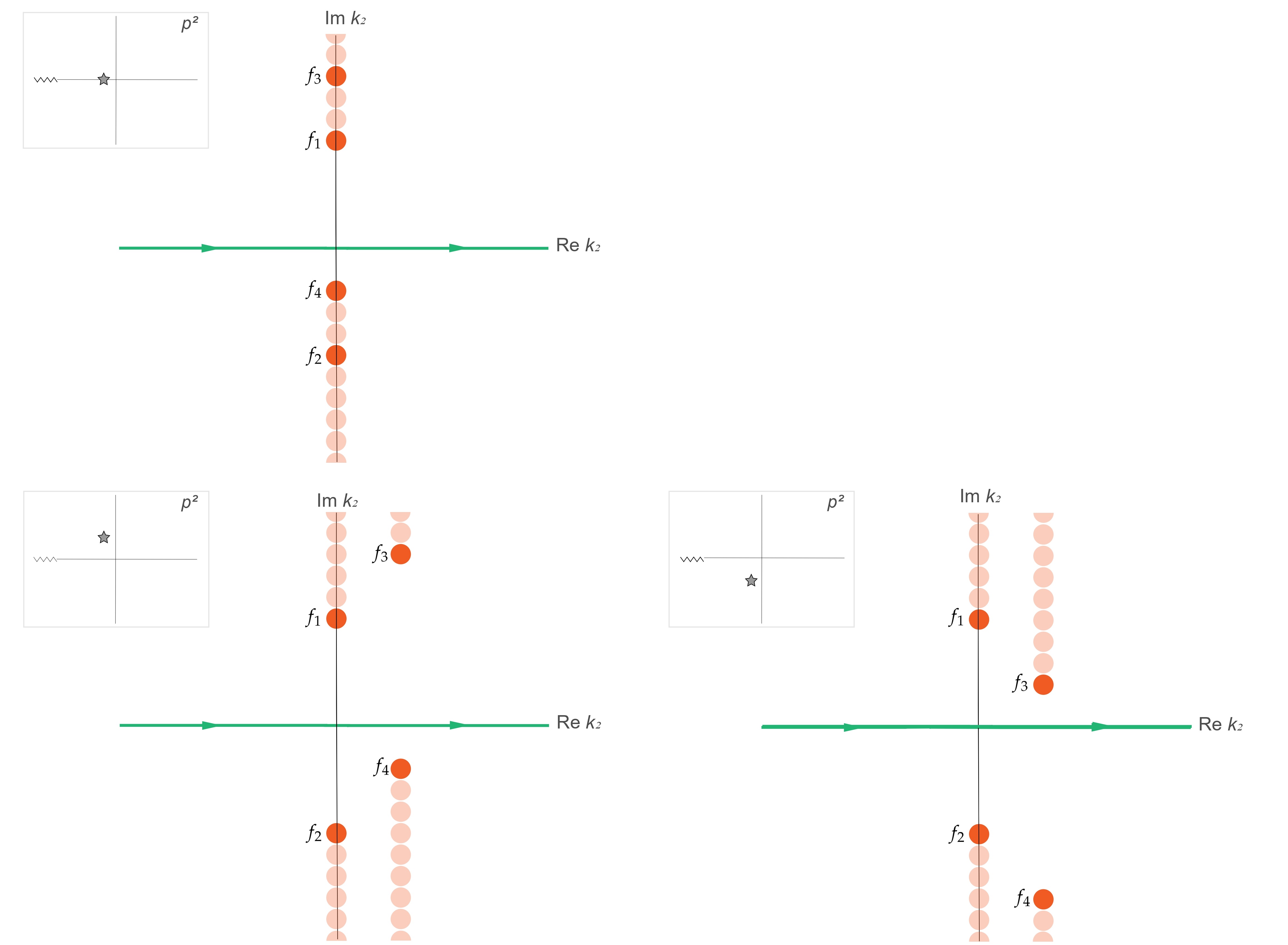}
	\caption{The $k_0$ integration path (green arrow) of the setting sun diagram in Euclidean spacetime $S_E(p^2)$ for three different values of the external momentum: $p^2 \in \mathbb{R}_-$ (top), $p^2 \in \mathbb{C}^+$ (bottom left) and  $p^2 \in \mathbb{C}^-$ (bottom right), as indicated in the inset. The orange dots indicate the poles $f_i$ for different values of $k$, with the darker dots indicating the poles for $k=0$.}
 \label{fig1}
\end{figure}

In Figure ~\ref{fig1} we show schematically the poles of the function \eqref{two2} in the complex $k_2$-plane. We show three examples: $p^2$ on the negative real axis and two complex conjugate values for $p^2$ above and below the negative real axis. We assume for now that $\vert\Im(p^2)\vert\gg \epsilon$ in the latter two cases to not interfere with the $+i \epsilon$ description. In these examples, $\Im(\sqrt{p^2})>-m$ so that the integrand of \eqref{two2} does not develop any poles as $k_2$ and $k$ vary along the real axis. In the case that $\Im(\sqrt{p^2})<-m$ direct evaluation is not possible and the integral requires an analytic continuation. The $k_2$-integral can now be performed by closing the integral on the upper-half plane and using Cauchy's residue theorem summing over the residues of the enclosed poles. In all four cases they are given by $R_1(k,p)$ and $R_3(k,p)$, and we have

\beq
S_E\left(p^2\right)&=&\frac{i}{2 \pi} \int dk  \left(R_1(k,\sqrt{p^2})+R_3(k,\sqrt{p^2})\right)\nonumber\\
&=& \frac{1}{2\pi} \int dk \frac{1}{\sqrt{k^2+m^2}(4k^2+4m^2+p^2)}\nonumber\\
&=& -\frac{i}{\pi} \frac{\arctan (\frac{i \sqrt{p^2}}{\sqrt{4m^2+p^2}})}{\sqrt{p^2} \sqrt{4m^2+p^2}}\label{tussendoor}
\eeq
and we can check that this is equal to the spectral representation from Eq.~\eqref{se} in the whole cut complex plane. One also sees that since $\arctan(x)/x$ is an even function, this function is single-valued on the negative real $p^2$ axis for $-4m^2<p^2<0$. We see that while the residues $R_1$ and $R_3$ each possess a branch cut on the negative $p^2$-axis, their sum does not for $p^2>-4m^2$.\\

We now want to investigate the relationship with the Minkowski formulation of the sunset diagram, in particular, we want to see if we can identify $S_E \left(p^2\right) =  S_M \left(-p^2\right)$. We have in two dimensions
\beq
S_M\left(-p^2\right)=\frac{1}{(2 \pi)^2}\int dk \int dk_0\frac{-i}{\left(k_0^2-k^2-m^2+i \epsilon \right)} \frac{1}{\left((k_0-\sqrt{-p^2})^2-k^2-m^2+i \epsilon \right)}. 
\label{fo5}
\eeq
We first need to clarify the relation between $\sqrt{-p^2}$ and $\sqrt{p^2}$ as defined above. We use the relation $\sqrt{z_1 z_2}=\sqrt{\left|z_1 z_2\right|} e^{i \operatorname{Arg}\left(z_1 z_2\right) / 2}$ from \cite{vanEgmond:2022nuo} with $z_1=-1$ and $z_2=p^2$ . This gives
\beq
\sqrt{-p^2}=\sqrt{\vert p^2 \vert}e^{i \text{Arg}(-p^2)/2}= \begin{cases} -i \sqrt{p^2}\,\,\, &\text{for}\,\,\, p^2 \in \mathbb{C}^+ \\ i \sqrt{p^2}\,\,\, &\text{for}\,\,\, p^2 \in \mathbb{C}^- \\ \sqrt{\vert p^2 \vert} \,\,\, &\text{for}\,\,\, p^2 \in \mathbb{R}^- \end{cases},
\label{lbl}
\eeq
where we used that $\text{Arg}(-p^2)=\text{Arg}(p^2)\pm \pi$ for $p^2 \in \mathbb{C}^{\mp}$. In this way, $\sqrt{-p^2}$ is well defined everywhere except on the positive real $p^2$-axis, and also $\overline{\sqrt{-p^2}}=\sqrt{-\overline{p^2}}$ for all $p^2 \in \mathbb{C}/\mathbb{R}_-$. The $k_0$-integrand has four $k$-dependent poles 
\beq
g_1=\sqrt{k^2+m^2}-i \epsilon, \,\,\,
g_2=-\sqrt{k^2+m^2}+i \epsilon , \,\,\,
g_3=\sqrt{k^2+m^2}-i \epsilon+\sqrt{-p^2}, \,\,\,
g_4=-\sqrt{k^2+m^2}+i \epsilon+\sqrt{-p^2}
\label{poles}
\eeq
with residues being respectively $R_4(k,i \sqrt{-p^2})$, $R_3(k,i \sqrt{-p^2})$, $R_2(k, i \sqrt{-p^2})$ and $R_1(k, i \sqrt{-p^2})$, using the definitions of Eq.~\eqref{ten}.\footnote{Unlike for the poles, at the level of the residues we can safely set $\epsilon \to 0^+$.}\\

\begin{figure}[!h]
	\includegraphics[width=\textwidth]{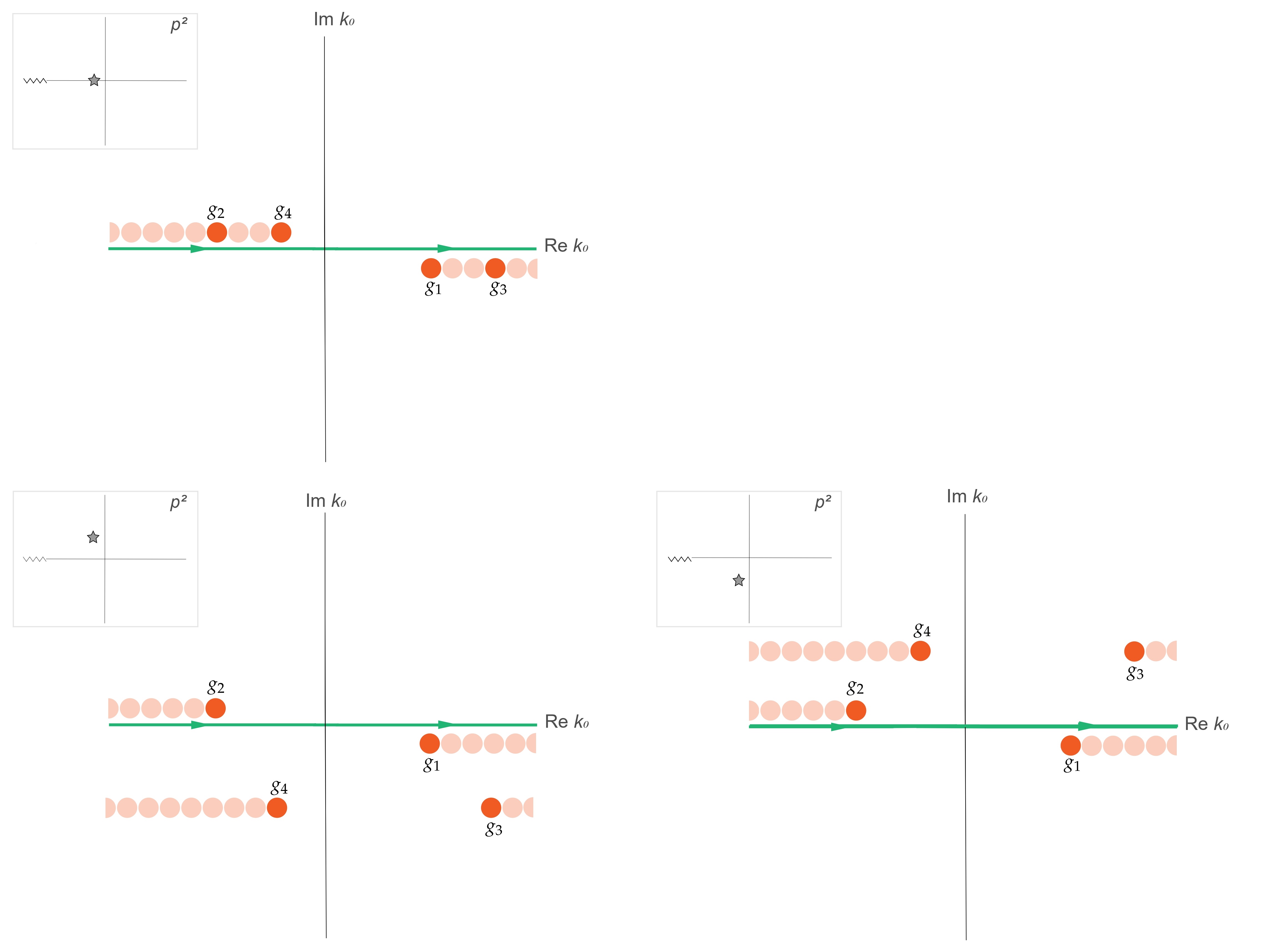}
	\caption{The $k_0$ integration path (green arrow) of the setting sun diagram in Minkowski spacetime $S_M(-p^2)$ for three different values of the external momentum: $p^2 \in \mathbb{R}_-$ (top), $p^2 \in \mathbb{C}^+$ (bottom left) and  $p^2 \in \mathbb{C}^-$ (bottom right), as indicated in the inset. The orange dots indicate the poles $g_i$ for different values of $k$, with the darker dots indicating the poles for $k=0$.}
 \label{fig2}
\end{figure}

In Figure \ref{fig2} the positions of the poles of the function \eqref{fo5} are made visible for the same three values of $p^2$ as in the Euclidean case. For $p^2$ on the negative real axis, closing the contour on the upper-half plane will enclose the poles $g_2$ and $g_4$. Using Cauchy's residue theorem, we then sum over $R_3(k,i\sqrt{-p^2})$ and $R_1(k,i\sqrt{-p^2})$. We then get Eq.~\eqref{tussendoor} so that $S_M\left(-p^2\right)= S_E(p^2)$ for $-4m^2<p^2<0$, as expected. On the other hand, for $p^2 \in \mathbb{C}^+$, closing the contour on the upper-half plane will only enclose $g_2$, so that
\beq
S_M(-p^2)= \frac{i}{2\pi} \int dk R_3 (k,i \sqrt{-p^2}) = S_E(p^2)-  \frac{i}{2\pi} \int dk R_1(k, i \sqrt{-p^2})
\label{12}
\eeq
Lastly, for $p^2 \in \mathbb{C}^-$ closing the contour on the upper-half plane will enclose the poles $g_2, g_3$ and $g_4$ so that
\beq
S_M(-p^2)&=& \frac{i}{2\pi} \int dk \left\{R_1 (k,i \sqrt{-p^2})+R_3 (k,i \sqrt{-p^2})+R_2 (k,i \sqrt{-p^2}) \right\}\nonumber\\
&=& S_E(p^2)+\frac{i}{2\pi} \int dk R_2(k,i\sqrt{-p^2})
\label{13}
\eeq
Using Eq.~\eqref{lbl} we then get for both $p^2 \in \mathbb{C}^+$ and $p^2 \in \mathbb{C}^-$ after performing the integral
\beq
S_M(-p^2)=\frac{1}{2} S_E(p^2)-\frac{i}{4 \sqrt{p^2} \sqrt{4m^2+p^2}},
\label{14}
\eeq
which is however neither well-defined for $p \in \mathbb{R}_-$ nor can it be analytically connected to the result $S_M\left(-p^2\right)= S_E(p^2)$ valid for $-4m^2<p^2<0$.

We have thus found that for three different choices ($p^2 \in \mathbb{C}^+, p^2 \in \mathbb{R}^-$ and $p^2 \in \mathbb{C}^+$) we find different expressions which are not connected through analytical continuation. This can be explained by the fact that the integrand in Eq.~\eqref{fo5} develops real poles as $p^2$ approaches the real negative axis, as can be seen in the upper graph of Figure \ref{fig1}. Moving $p^2$ into the upper half-plane, the poles $g_3$ and $g_4$ as defined in \eqref{poles} will move downwards until at some point, $g_4$ reaches the real $k_0$-axis. At this point, the contour integral will have to be deformed to go just below the real axis. On the other hand, starting from $p^2 \in \mathbb{C}^+$ as in the bottom right graph of Figure \ref{fig1} and moving $p^2$ downwards, i.e. moving the poles $g_3$ and $g_4$ upwards, when the pole $g_4$ reaches the real $k_0$-axis we will have to deform the contour to go just above the real axis. Using 
\beq
\lim_{\epsilon \to 0^+} \int_{-\infty}^{\infty} f(x \pm i\epsilon)\, dx 
= \operatorname{PV} \int_{-\infty}^{\infty} f(x)\, dx \mp i\pi \cdot \operatorname{Res}(f, x_0)
\eeq
where PV stands for Cauchy's Principal Value, we see that the difference between the pole $g_4$ approaching the real axis from above and below is exactly given by the extra term in the second line of Eq.~\eqref{12}. In the same way, moving $p^2$ away from the negative real axis into the lower half-plane, at some point $g_3$ will reach the real $k_0$-axis from below and the contour integral will have to be deformed to go just above the real $k_0$-axis. If instead we start with $p^2 \in \mathbb{C}^-$ the pole $g_3$ reaches the real $k_0$-axis from above and the deformed contour integral will go just below the real $k_0$-axis. The difference is given by the extra term in the second line of \eqref{13}. In other words, the costs of moving the poles over the real integration axis is given by eqs.~\eqref{12} and \eqref{13}. Note that since neither half-plane in $p^2$-space is smoothly connected to the real axis, the Schwarz reflection principle, which would imply $\overline{S_M(p^2)}=S_M(\overline{p^2})$, cannot be invoked by a unique analytical continuation. 

The above result is in contradiction with the spectral representation from Eq.~\eqref{se}, which predicts that $S_M(-p^2)=S_E(p^2)$ in the whole cut complex plane. We therefore have to conclude that for the setting sun diagram, performing the integral over the internal momenta with $\textit{a priori}$ complex external momenta does not lead to the expected results. \\

\section{Wick rotation}
There is another way we can see that the analytic continuation between the Minkowskian and Euclidean expression of the setting sun fails when working with $\textit{a priori}$ complex external momenta, which is through the usual method of the Wick rotation. Let us start with the usual case of $-4m^2<p^2<0$ so that in the Minkowskian case the poles are located just above and below the real $k_0$-axis, as in the top left graph in Figure \ref{fig2}. Making a counterclockwise Wick rotation to the imaginary axis, we get
\beq
S_M\left(-p^2\right)=\frac{1}{(2 \pi)^2}\int dk \int_{-i \infty}^{i \infty} dk_0\frac{-i}{\left(k_0^2-k^2-m^2+i \epsilon \right)} \frac{1}{\left((k_0-\sqrt{-p^2})^2-k^2-m^2+i \epsilon \right)}. 
\eeq
and redefining $k_0=-i k_2$, we have

\beq
S_M\left(-p^2\right)=\frac{1}{(2 \pi)^2}\int dk \int_{\infty}^{\infty} dk_0\frac{1}{\left(k_2^2+k^2+m^2 \right)} \frac{1}{\left((k_2+i \sqrt{-p^2})^2+k^2+m^2\right)},
\label{ll}
\eeq
where we have safely set $\epsilon \to 0^+$. Equation \eqref{ll} is equal to $S_E(p^2)$ with $\sqrt{p^2}$ replaced by $i \sqrt{-p^2}$. The function $i \sqrt{-p^2} = \pm \sqrt{p^2}$ is not uniquely defined for  $(-4m^2<)p^2<0$, but as we have seen above in Eq.~\eqref{tussendoor}, $S_E(p^2)$ eventually is an even function of $\sqrt{p^2}$, so either sign will give the same expression. We thus find that $S_M(-p^2)=S_E(p^2)$ through Wick rotation.\\

Next, we try to follow the same procedure for $p^2 \in \mathbb{C}/\mathbb{R}$, that is, we try to Wick rotate from the lower two graphs in Figure \ref{fig2} to the lower two graphs in Figure \ref{fig2}. We can see that there is no Wick rotation possible that does not pick up poles. In fact, the poles that are picked up for both $p^2$ in the upper-half plane and the lower-half plane correspond to the residue $R_1(k,\sqrt{p^2})$, so that we get exactly the same identification as in Eq.~\eqref{14}. This means that also from a Wick rotation, one would not get $S_M(-p^2)=S_E(p^2)$ corresponding to the results from the spectral representation \eqref{se}.

All in all, we have illustrated that directly computing the Minkowskian Feynman integral \eqref{two2} for the most general complex momentum configuration leads to undesired results, not compatible with the generally accepted constraints on quantum field theory correlators.

\section{Alternative formulation of the Minkowski propagator}\label{sectminkalt}

\begin{figure}[!h]
	\includegraphics[width=\textwidth]{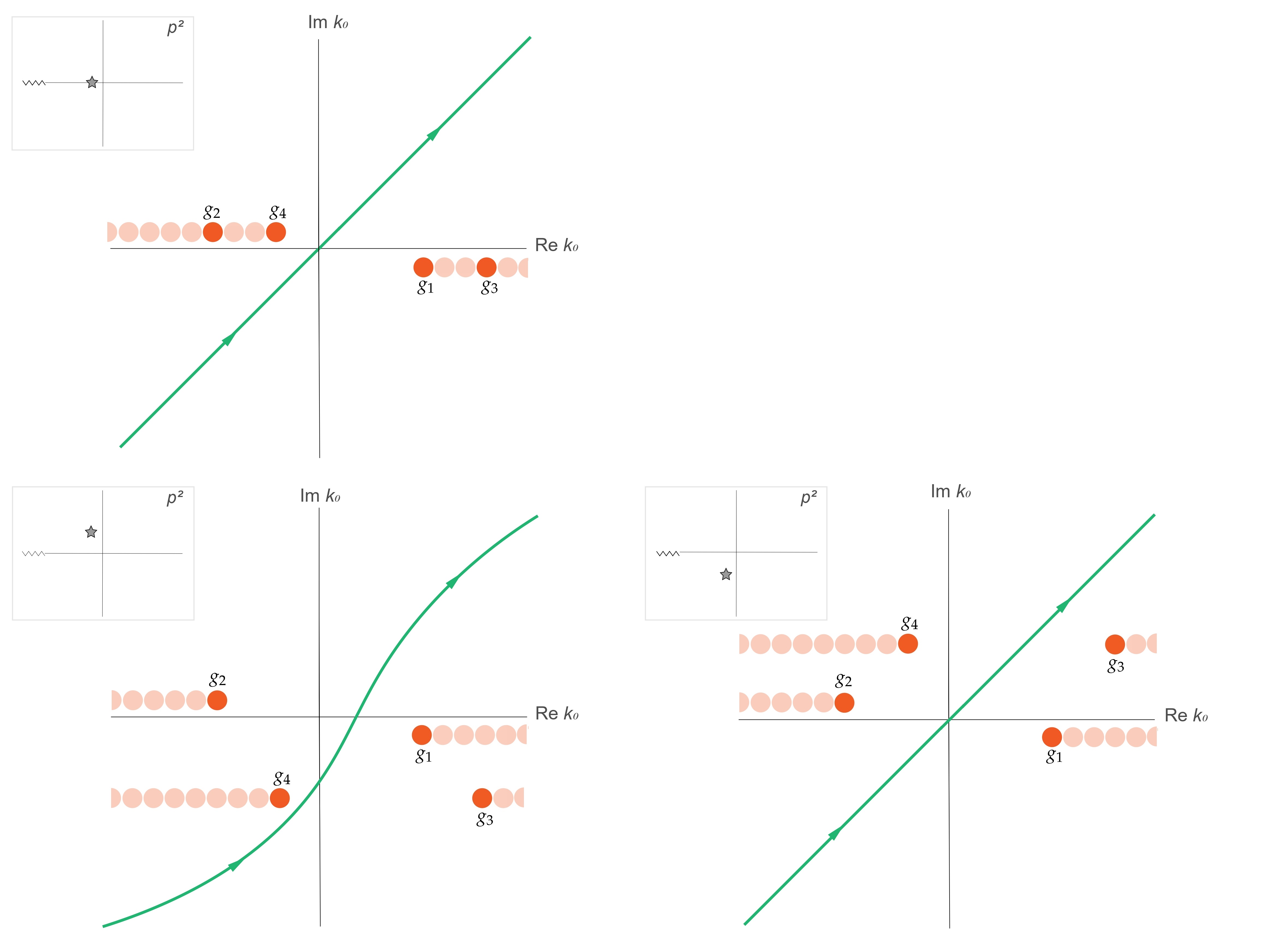}
	\caption{The $k_0$ integration path (green arrow) of the setting sun diagram in Minkowski spacetime $S_M(-p^2)$ in the alternative formulation of Eq.~\eqref{fo3} for three different values of the external momentum: $p^2 \in \mathbb{R}_-$ (top), $p^2 \in \mathbb{C}^+$ (bottom left) and  $p^2 \in \mathbb{C}^-$ (bottom right), as indicated in the inset. The orange dots indicate the poles for different values of $k$, with the darker dots indicating the poles for $k=0$.}
 \label{fig3}
\end{figure}

We now discuss a proposal given in \cite{Eichmann:2019dts} which seems to resolve the discrepancy in the relation between the Euclidean and Minkowskian picture found through the K\"all\'en-Lehmann spectral representation and the direct integration as performed in the last section. 

According to \cite{Eichmann:2019dts}, the correct path integral in Minkowski space must read
\beq
S_M\left(p^2\right)=\frac{1}{(2 \pi)^2} \int d k \int_{-\infty(1+i \epsilon)}^{\infty(1+i \epsilon)} d k_0\frac{1}{\left(k_0^2-k^2-m^2\right)} \frac{1}{\left((k_0-\sqrt{p^2})^2-k^2-m^2 \right)},
\label{fo3}
\eeq
rather than the formulation in Eq.~\eqref{fo2}. This is based on the following fact, rooted in the discussion in for example \cite{Peskin:1995ev}. Consider $\left|\Omega \right \rangle$ the ground state of a free Hamiltonian $H_0$, and $\ket{0}$ that of an interaction-included $H$, where as usual it is assumed these 2 ground states have a nonzero overlap. Considering the exact time evolution operator's action on the non-interacting ground state,
\begin{equation}\label{time0}
    e^{-iHT}\left|\Omega \right \rangle= e^{-iE_0T}\left|0 \right \rangle\braket{0|\Omega}+ \sum_{n\neq0} e^{-iE_nT}\left|n \right \rangle\braket{n|\Omega}
\end{equation}
with $\left|n \right \rangle$ a ``discrete'' notation for the spectrum of $H$.  We want to isolate the exact ground state from the above sum, which can be done by realizing that in a well-defined field theory, the energy eigenvalues will be real with $E_n>E_0$, so that
\begin{equation}\label{time1}
    \ket{0}=\frac{1}{\braket{0|\Omega}}\lim_{T\to\infty(1-i\epsilon)} e^{-i(H-E_0)T}\ket{\Omega}
\end{equation}
and it is this $\ket{0}$ that is used to construct time-ordered field expectation values in the interacting theory. Clearly, the tiny imaginary part serves to kill off all unnecessary terms, viz.
\begin{equation}\label{time2}
    \lim_{T\to\infty(1-i\epsilon)} e^{-i(E_n-E_0)T}=0\,,\qquad \forall n\neq 0
\end{equation}
This amounts in momentum space to  $k_0 \to \pm\infty(1+i \epsilon)$ and thus the integral in Eq.~\eqref{fo3} seems to be the basic one rather than the one of Eq.~\eqref{fo2}. For real $p^2$, these formulations are actually equivalent, as can be seen by substituting $k_0 \to k_0(1+i \epsilon)$ in \eqref{fo3}. For complex $p^2$, however, this depends on the interpretation of $\epsilon \infty$: if we consider that $k_0 (i \epsilon)$ gives a small imaginary value for arbitrarily large $k_0$, as we have assumed in the last section, we can identify \eqref{fo2} and \eqref{fo3}. If, on the other hand, we consider that $k_0 (i \epsilon)$ becomes arbitrarily large for arbitrarily large $k_0$, there will always be a real pole since $k_0 \epsilon$ will cancel with the complex part of $\sqrt{p^2}$ at some point, while in the formulation of Eq.~\eqref{fo2} the poles are always located outside the real axis. Starting from the integral \eqref{fo3} and following the premise that $k_0 \epsilon$ indeed can become arbitrarily large for arbitrarily large $k_0$, we can find the original relation between $S_E(p^2)$ and $S_M(-p^2)$, since now the Minkowski integration path is not identified with the imaginary $k_0$-axis, but rather with the path going from the lower right corner to the upper-left corner. From Figure~\ref{fig3} we see immediately that we will not have any problem smoothly connecting $p^2$ in the three different scenarios ($p^2 \in \mathbb{C}^+$, $p^2 \in \mathbb{R}^-$ and $p^2 \in \mathbb{C}^-$) since the integration line is not obstructing the poles from moving up and down. 
No matter where the poles are located, one can always find such a path that can then be rotated to the imaginary axis without crossing any poles and the identification $S_M(-p^2)=S_E(p^2)$ is direct for all $p \in \mathbb{C}$.

There can, however, be made some remarks about the use of \eqref{fo3} as following from the correct path integral in Minkowski space. The first is the use of $\epsilon \infty \to \infty$. This is ambiguous at least, since the equivalence of the Minkowski and Euclidean picture is based on setting $\epsilon \to 0^+$ after rotation to the imaginary axis. Secondly, the use of $T \to \infty (1+ i \epsilon)$ is used to single out the ground state, but this does not work in cases where e.g.~complex conjugate masses are included in more general quantum field theories, such as the Curci-Ferrari effective model, based on \cite{Curci:1976bt}, which received a lot of renewed attention during the last decade as an effective description of non-perturbative Yang-Mills Green's functions or thermodynamics, see the recent review \cite{Pelaez:2021tpq}. In \cite{Kondo:2019rpa}, it was, however, noted that starting from one loop onward, the model develops complex gluon poles as well. Complex poles are also akin to the Gribov-Zwanziger formalism, a way to deal with the gauge copies \cite{Gribov:1977wm,Zwanziger:1988jt,Dudal:2008sp,Baulieu:2009ha}. 

Assume indeed that for some $n=n_\pm$, $E_{n_\pm} = a\pm ib$ with $b\neq0$ appears, then the limit \eqref{time2} can pick up an exploding ($\epsilon$-independent) factor due to the imaginary parts in $E_{n_\pm}$. This could be avoided if the $\sum_n$ could be restricted to physically meaningful states $\ket{n}$. To be more precise, we would have to find a resolution of the identity in a suitably defined physical subspace with only real $E_n>E_0$, built from the unphysical states. This is, however, an entirely different discussion and as of now, we are unaware of any such concrete procedure in e.g.~the Curci-Ferrari or Gribov-Zwanziger models.

Finally, it is not clear how the integration path in Eq.~\eqref{fo3} is to be drawn in the case that $p^2$ goes towards an infinite complex number, something which becomes relevant in the case of branch cuts in $p^2$-space. Indeed, if at some order in perturbation theory a pair of Euclidean complex pole masses appear, at higher order these poles will generate branch points arbitrarily deep into the complex $p^2$-plane, order per order. 

\section{Conclusion}\label{sectcon}
In this short note, we have investigated various possibilities of connecting the two-dimensional setting sun diagram in Euclidean spacetime with its equivalent in Minkowskian spacetime in the case of generic complex external momenta. Such endeavor will only become more important in follow-up work when one must face the consequence of working with Green's function that have complex poles to begin with. For now, sticking to two equal tree-level real masses $m$, we know from complex analysis and through the K\"all\'en-Lehmann spectral representation that the Euclidean and Minkowskian expressions can be related with the same identification for both real and complex $p^2$. However, this does not match with the outcome of a priori starting from an Feynman integral with complex external momentum, as we have shown, and one should therefore not start with a complex value of $p^2$ inside the integral, but first perform the integral for real $p^2$ and properly extend the result to complex $p^2$ to match with the spectral representation.  Such considerations become even more important in the often elaborate numerical approaches to calculations of e.g.~quark and gluon propagators. For a recent example, see \cite{Oribe:2025ezp} where the leading order renormalization group improved Euclidean and Minkowskian Curci-Ferrari propagators were considered in a specifically constructed scheme, but no simple analytic continuation or spectral representation was unveiled as of now due to the complications involved.

We have discussed the alternative formulation of the Minkowskian setting sun diagram as proposed in \cite{Eichmann:2019dts}, which seems to connect directly with the desired K\"all\'en-Lehmann integral representation. However, we also pointed out some potential caveats with the interpretation of $\infty \times \epsilon$ or for singularities at arbitrarily large momenta, something that will become most relevant when there are complex poles and cuts.

Finally, we want to mention here the work of \cite{Siringo:2022dzm} on the analytical continuation of the gluon propagator in presence of complex mass poles, as they appear in the loop expansion of the Curci-Ferrari model, see also \cite{Kondo:2019rpa,Hayashi:2020few,Hayashi:2021jju}. The there discussed analytical continuation will then lead to opposite signs for the pole structures in the corresponding Minkowski propagator, similar to our findings in Section~\ref{22}. Henceforth, a generalized K\"all\`en-Lehmann spectral density function was constructed. This can become an important clue as how to deal with integrals where one does not have a standard spectral density function to benchmark results, such as for the setting sun diagram with complex conjugate tree-level masses, to which we will return in forthcoming work. 

\section*{Acknowledgments}
The work of D.~Dudal was supported by KU Leuven IF project C14/21/087.  D.M. van Egmond was supported by FAPESP grants 2023/03722-9 and 2024/10642-4 and would like to thank KU Leuven for their hospitality during the final stages of this work. The work of G.~Krein was supported in part by Conselho Nacional de Desenvolvimento Cient\'ifico e Tecnol\'ogico (CNPq) under Grant No. 309262/2019-4, and by Funda\c{c}\~ao de Amparo \`a Pesquisa do Estado de S\~ao Paulo (FAPESP) under Grant No. 2018/25225-9. We are grateful for insightful discussions with G.~Comitini, P.M.~Lo and U.~Reinosa.

\bibliographystyle{apsrev4-2}

\end{document}